\documentclass[preprint,showpacs,amsmath,amssymb,prd]{revtex4}
\usepackage{bm}     
\usepackage{psfig}  
\newcommand{\ul}{}
\newcommand{\ull}{}
\newcommand{\dedx}{dE/dx}
\newcommand{\BR}{\mathcal{B}}

\newcommand{\eff}{\varepsilon}

\newcommand{\gev}{\,\mbox{GeV}}

\newcommand{\ra}{\rightarrow}

\newcommand{\psp}{\psi^\prime}

\newcommand{\jpsi}{J/\psi}
\newcommand{\pspp}{\psi^{\prime\prime}}
\newcommand{\psppto}{\psi^{\prime\prime}\to}
\newcommand{\EE}{e^+e^-}
\newcommand{\EETO}{e^+e^-\to}
\newcommand{\pip}{\pi^+}
\newcommand{\pim}{\pi^-}
\newcommand{\piz}{\pi^0}
\newcommand{\threepi}{\pi^+\pi^-\pi^0}

\newcommand{\rhopi}{\rho\pi}
\newcommand{\bfg}{\begin{figure}}
\newcommand{\efg}{\end{figure}}
\newcommand{\bitm}{\begin{itemize}}
\newcommand{\eitm}{\end{itemize}}
\newcommand{\bnum}{\begin{enumerate}}
\newcommand{\enum}{\end{enumerate}}
\newcommand{\btbl}{\begin{table}}
\newcommand{\etbl}{\end{table}}
\newcommand{\btbu}{\begin{tabular}}
\newcommand{\etbu}{\end{tabular}}
\begin{document}


\title{\boldmath Search for \ull{$\psi(3770)\ra\rhopi$} at the BESII
detector at the Beijing Electron-Positron Collider}
\author{ M.~Ablikim$^{1}$,      J.~Z.~Bai$^{1}$,            Y.~Ban$^{11}$,
  J.~G.~Bian$^{1}$,      X.~Cai$^{1}$,               H.~F.~Chen$^{15}$,
  H.~S.~Chen$^{1}$,      H.~X.~Chen$^{1}$,           J.~C.~Chen$^{1}$,
  Jin~Chen$^{1}$
% ...
  \\(BES Collaboration)\\}

\affiliation{
  $^{1}$ Institute of High Energy Physics, Beijing 100049, People's Republic of China\\
% ...
$^{11}$ Peking University, Beijing 100871, People's Republic of China\\
% ...
$^{15}$ University of Science and Technology of China, Hefei 230026, People's Republic of China\\
% ...
}

\date{\today}

\begin{abstract}
Non-$D\bar{D}$ decay \ull{$\psi(3770)\ra \rhopi$} is searched for
using a data
sample of $(17.3\pm 0.5)~pb^{-1}$ taken at the center-of-mass
energy of 3.773 GeV by the BESII detector at the BEPC. No
$\rhopi$ signal is observed, and the upper limit of the cross
section is measured to be $\sigma(\EETO \rhopi)<6.0~pb$
at 90\% C. L. Considering the interference between the continuum
amplitude and the \ull{$\psi(3770)$} resonance amplitude,
the branching fraction of \ull{$\psi(3770)$} decays to $\rho\pi$ is
determined to be
\ull{$\BR(\psi(3770)\ra\rho\pi)\in(6.0\times10^{-6},~2.4\times10^{-3})$}
at 90\% C. L. This is in agreement with the prediction of the $S$-
and $D$-wave mixing scheme of the charmonium states for solving the
``$\rhopi$ puzzle'' between $\jpsi$ and \ull{$\psi(2S)$} decays.
\end{abstract}

\pacs{13.25.Gv, 12.38.Qk, 14.40.Gx}

\maketitle

\section{Introduction}

\subsection{\boldmath ``$\rhopi$ puzzle'' and \ull{$\psi(3770)\ra \rhopi$}}

Perturbative QCD (pQCD) predicts that the decays of $\jpsi$ and
$\psi(2S)$ (shortened as $\psp$ below) into light hadrons are
dominated by the annihilation of $c\bar{c}$ into three gluons,
with widths proportional to the square of the wave function at the
origin~\cite{appelquist}. This yields the pQCD ``12\% rule'', that
is
\begin{eqnarray}
Q_h &=&\frac{{\cal B}_{\psp \ra h}}{{\cal B}_{\jpsi \ra h}}
=\frac{{\cal B}_{\psp \ra \EE}}{{\cal B}_{\jpsi \ra \EE}} \approx
 12\%.
\label{qcdrule}
\end{eqnarray}

The violation of the above rule was first observed in $\rhopi$ and
$K^{*+}K^-+c.c.$ modes by Mark II~\cite{mk2}, known as the {\it
$\rhopi$ puzzle}. Since then BES and CLEO-c have measured many
two-body decay modes of $\psp$, among which some obey the 12\%
rule while others violate it~\cite{guyf,cleocvp}. There have been
many theoretical efforts trying to solve the
puzzle~\cite{puzzletheory}. A recent one is the $S$- and $D$-wave
charmonia mixing model proposed by Rosner~\cite{rosnersd}. In this
scheme, the mixing of the $\psi (2^3 S_1)$ and $\psi (1^3 D_1)$
states is in such a way that there is almost a complete
cancellation of the decay amplitude of $\psp \rightarrow \rhopi$,
which instead shows up as an enhanced decay mode of
\ull{$\psi(3770)$ (shortened as $\pspp$ below)}. A study shows
that in $\EE$ experiments with $\BR(\pspp \rightarrow\rhopi)$
predicted by the $S$- and $D$-wave mixing in Ref.~\cite{rosnersd},
the destructive interference between the three-gluon decay
amplitude of the $\pspp$ resonance and the continuum one-photon
amplitude leads to a very small $\rho\pi$ cross section at
\ull{the $\pspp$ peak}~\cite{wympspp}, which is in agreement with
the unpublished upper limit of the $\rhopi$ cross section at the
$\pspp$ peak by Mark~III~\cite{mk3}. Using \ull{a larger} data
sample to further study this channel will shed light on the
understanding of the ``$\rhopi$ puzzle'' and the interference
pattern between the resonance and the continuum amplitudes.

\subsection{Data samples and detector}

The data used for this analysis are taken with the BESII detector
at the BEPC storage ring at \ul{the center-of-mass energy} of
$3.773~\gev$. The integral luminosity of the data sample is
$(17.3\pm 0.5)~pb^{-1}$ as measured using large angle Bhabha
events. To study the continuum process, BESII also collected
$(6.42\pm 0.24)~pb^{-1}$ data at $\sqrt{s}=3.65~\gev$~\cite{con_da}.
We analyze these two data samples simultaneously to get the numbers
of events at \ull{the $\pspp$ peak} and at the continuum.

BESII is a conventional solenoidal magnet detector that is
described in detail in Refs.~\cite{bes,bes2}. A 12-layer vertex
chamber (VC) surrounding the beam pipe provides trigger
information. A 40-layer main drift chamber (MDC), located
radially outside the VC, provides trajectory and energy loss
($\dedx$) information for tracks over $85\%$ of the
total solid angle.  The momentum resolution is
$\sigma _p/p = 0.017 \sqrt{1+p^2}$ ($p$ in $\hbox{\rm GeV}/c$),
and the $\dedx$ resolution for hadron tracks is $\sim 8\%$.
An array of 48 scintillation counters surrounding the MDC  measures
the time-of-flight (TOF) of tracks with a resolution of
$\sim 200$ ps for hadrons.  Radially outside the TOF system is a 12
radiation length, lead-gas barrel shower counter (BSC).  This
measures the energies of electrons and photons over $\sim 80\%$ of the
total solid angle with an energy resolution of
$\sigma_E/E=22\%/\sqrt{E}$ ($E$ in GeV).  Outside of the solenoidal coil,
which provides a 0.4~Tesla magnetic field over the tracking volume,
is an iron flux return that is instrumented with
three double layers of  counters that
identify muons of momentum greater than 0.5~GeV/$c$.

\subsection{Monte Carlo Simulation}

A Monte Carlo simulation is used for the determinations of the
mass resolutions and the detection efficiencies.  This program
(SIMBES), which is Geant3 based, simulates the detector response,
including the interactions of secondary particles with the
detector material.  Reasonable agreement between data and Monte
Carlo simulation has been observed in various channels tested~\cite{simbes},
including $\EETO(\gamma) e^+e^-$, $\EETO (\gamma)\mu^+\mu^-$,
$\jpsi\ra p\bar{p}$ and $\psp \ra \jpsi \pip\pim, \, \jpsi \ra
\ell^+\ell^-$ $(\ell=e,\mu)$.

For the signal process, $\EE\ra \rhopi$, the Monte Carlo events
are generated \ull{with an} angular distribution of
$\sin^2\theta_1(1+\cos^2\theta_v+\sin^2\theta_v\cos(2\phi_1))$,
where $\theta_v$ is the angle between \ull{the $\rho$} and the
positron direction, and $\theta_1$ and $\phi_1$ are the polar and
azimuthal angles of the pion in the $\rho$ helicity frame. A Monte
Carlo sample of $\psppto \threepi$ is also generated with the same
angular distribution, \ull{taking into account the available
phase-space of the two-pion system}. The generators include the
effects of initial state radiation~(ISR), and the $\rhopi$ or
$\threepi$ form factor varies as a function of $s$ ($s$ denotes
the square of the center-of-mass energy), where a $1/s$ dependence
is assumed. The generator PPCON is also used to study the
interference between the continuum and resonant $\pspp$
amplitudes. In this generator, the branching fraction of
$\pspp\ra\rho\pi$ can be set to be any number between 0 and 1, the
relative phase between \ull{the $\pspp$} strong and
electromagnetic decay amplitudes can be set to be any possible
value from $-180^\circ$ to $180^\circ$ and the measurement of
$\sigma^{Born}(e^+e^-\ra\rho\pi)$ at $\sqrt{s}=3.67~\gev$ measured
by CLEO-c~\cite{cleocvp} is used to normalize the contribution of
the continuum cross section at \ull{the $\pspp$ peak}.

Monte Carlo samples of Bhabha, dimuon, $D\bar{D}$ and inclusive
hadronic events generated with Lundcrm~\cite{lundcharm} are used
for the background study.

\section{Event selection}

The final states of the study include two charged pions and one
neutral pion which is reconstructed from two photons. Event
selection includes photon identification and charged particle
identification.

A neutral cluster is considered to be a photon candidate when the
deposited energy in the BSC is greater than 80~MeV, the angle
between the nearest charged track and the cluster is greater than
$16^{\circ}$, and the angle between the two nearest photons is
larger than $7^{\circ}$. The first hit of the cluster is in the
beginning 6 radiation lengths, and the angle between the cluster
development direction in the BSC and the photon emission direction
must be less than $37^{\circ}$. The number of photon candidates
after the above selection is required to be two.

For each charged track, the TOF and $\dedx$ measurements are used
to calculate $\chi^2$ values and the corresponding confidence
levels to the hypotheses that the particle is a pion, a kaon or a
proton ($Prob_\pi$, $Prob_K$, $Prob_p$). A track is considered to
be a pion when the confidence level of the pion hypothesis is
greater than the confidence levels of the kaon and proton
hypotheses. At least one charged track is required to be
identified as a pion.

\begin{figure}[htbp]
\centerline{\hbox{\psfig{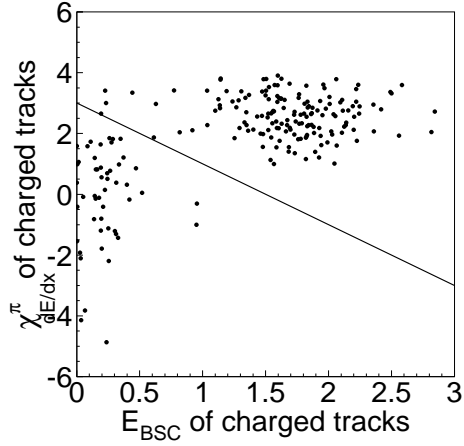}}}
\caption{Scatter plot of the $dE/dx$ separation from the pion
hypothesis versus the deposited energy in BSC of the charged
track. The cluster at the top right corner is radiative Bhabha
\ul{electrons}. The events above the straight line are removed as Bhabha
candidates.} \label{dedxbsc_pp}
\end{figure}

For the decay channel of interest, the candidate events must
satisfy the following selection criteria:
\bnum
     \item An event is required to have only two oppositely charged tracks in
     the MDC, each with a good helix fit.
     The closest approach of the track to the
     interaction point is required to be within 2~cm in the transverse
     plane and within 20~cm in the beam direction,
     and the transverse momentum
     $P_{xy}>0.06~\hbox{GeV/c}$ is used to remove the beam associated
     background.
     \item  A four-constraint kinematic fit is performed under the
  hypothesis $\EE\ra\gamma\gamma\pip\pim$, and the confidence level of
  the fit is required to be greater than 1\%. A four-constraint
  kinematic fit is also performed under the hypothesis of
  $\EE\ra\gamma\gamma K^+K^-$, and
  $\chi^2_{\gamma\gamma\pi\pi}<\chi^2_{\gamma\gamma KK}$ is required
  to remove \ull{the $K^+K^-\piz$} events.
     \item To remove the di-muon background and the backgrounds
      produced by the ISR \ull{process $\EE\ra\gamma\psp$},
      with $\psp\ra\hbox{neutral}+\jpsi$, $\jpsi\ra\mu^+\mu^-$,
      \ul{two tracks should be in the} $|\cos\theta|<0.80$ region
      ($\theta$ is the polar angle of the track in MDC)
      and at least one track is required to be in
      the coverage of muon counter,
     \ul{in which $N^{hit}_{\pi^+}+N^{hit}_{\pi^-}<3$ is
     required}. Here, $N^{hit}$ is the number of muon counter layers with matched hits
     and ranges from 0 to 3, indicating not a muon~(0), \ul{a weakly~(1),
     medianly~(2)}, or strongly~(3) identified muon track~\cite{muid}.
     \item After the four-constraint kinematic fit, the energy of
     the higher momentum photon candidate is required to be less than
     $1.5~\gev$ to remove the $\rho^0(770)$ background produced by ISR.
\item After the above selection, the radiative Bhabha background can
still be seen clearly from Figure~\ref{dedxbsc_pp}, where the
$dE/dx$ separation from pion hypothesis ($\chi^\pi_{dE/dx}$) and
the energy deposited in the BSC of the charged track ($E_{BSC}$)
are shown. The cluster of events in the top right corner are
electron tracks and can be removed by \ull{requiring
$\chi^\pi_{dE/dx}<-2E_{BSC}+3$} with high efficiency for the
signal events.
\enum

\begin{figure}[htbp]
\centerline{ \hbox{\psfig{file=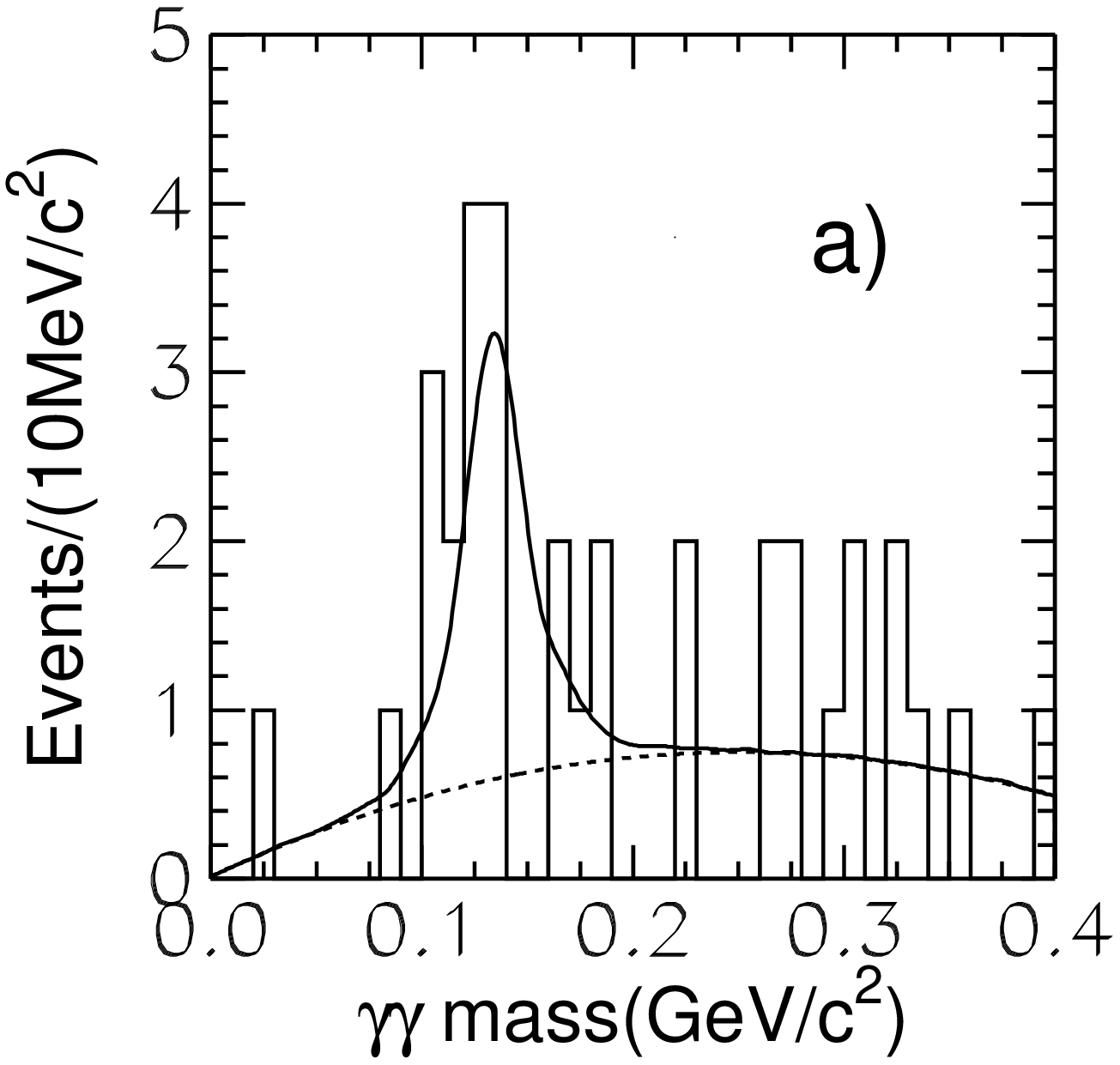,width=4.5cm}}
\hbox{\psfig{file=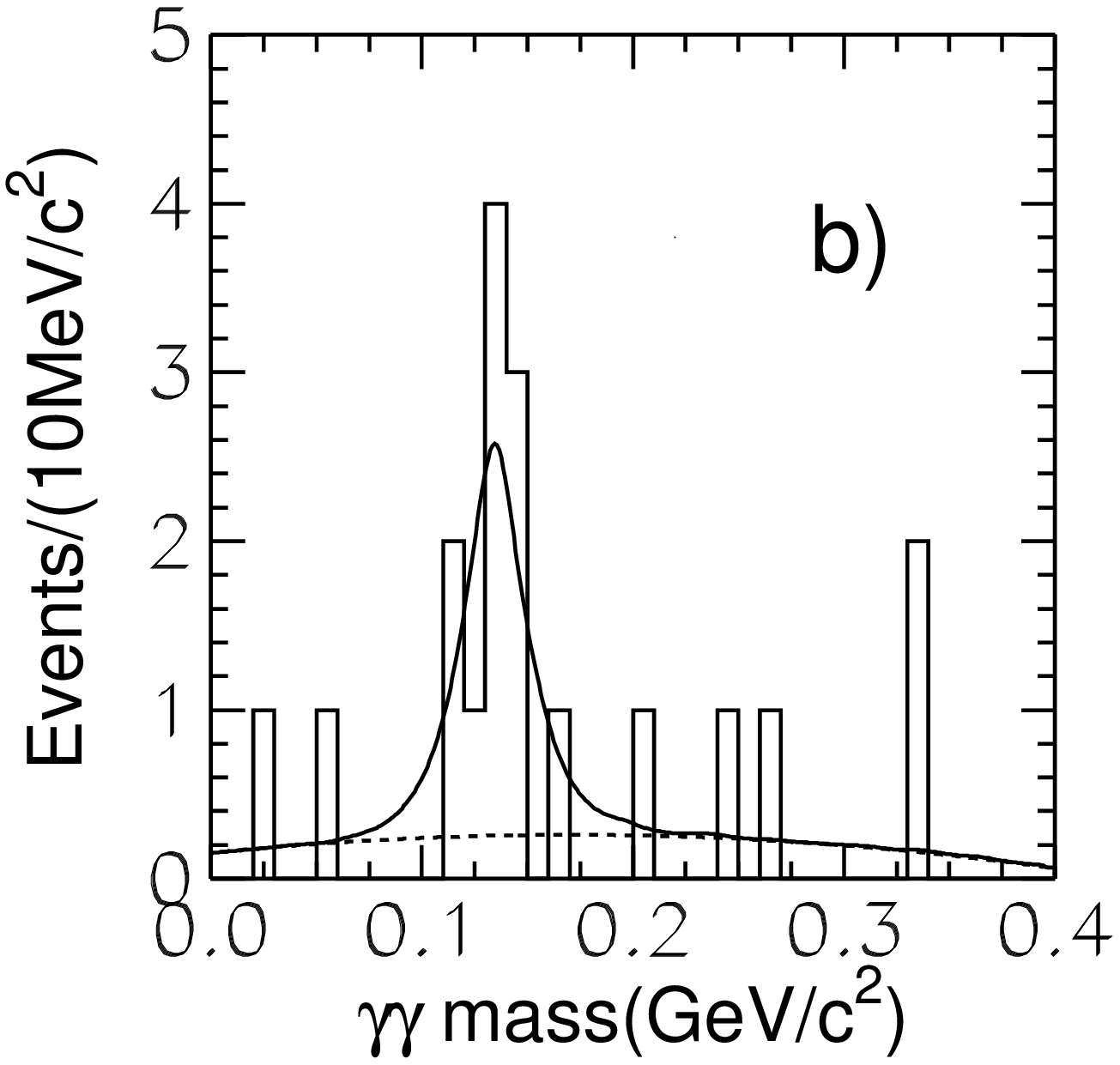,width=4.5cm}}       }
\caption{Invariant mass distributions of \ull{the two} photons
after final selection for (a) $\pspp$ data sample, and \ull{for
(b)} $\sqrt{s}=3.65~\hbox{GeV}$ data. The histograms are data, and
the curves show the best fits.} \label{mggfit}
\end{figure}

After applying all of the above selection criteria, the invariant
mass distributions of \ull{the two} photons after \ull{the
kinematic} fit are shown in Figures~\ref{mggfit}a and
\ref{mggfit}b for \ull{the $\pspp$} and \ull{the continuum} data
samples, respectively. It can be seen that there are $\piz$
signals in the $\pspp$ and $\sqrt{s}=3.65~\hbox{GeV}$ data
samples.

\ull{Analyses using the Monte Carlo samples of Bhabha, dimuon,
$D\bar{D}$, inclusive hadronic} \ull{events and ISR production of
$\jpsi$ and $\psp$ decays show that the background contaminations
to the} \ull{two-photon invariant mass} \ull{spectra are small or
in a random distribution.}

\section{\ull{Fits to the invariant mass spectra}}

The invariant mass spectra of \ull{the two} photons are fitted
with the Monte Carlo simulated $\piz$ invariant mass distribution
(where the $\piz$ mass is fixed to the PDG value~\cite{pdg} in the
simulation) for the signal and a 2nd-order polynomial for the
background. The fits yield $11.4\pm 4.7$ and $10.0\pm3.8$ $\piz$s
for \ull{the $\pspp$} and \ull{the $\sqrt{s}=3.65$~GeV} data
samples respectively, and the corresponding signal significance
are $3.1\sigma$ and $4.6\sigma$. The fit results are shown in
Figures~\ref{mggfit}a and \ref{mggfit}b. The efficiency of
detecting $\EETO \threepi$ at $\pspp$ is $(7.65 \pm 0.12)\%$ and
that at $\sqrt{s}=3.65$~GeV is $(7.88\pm0.12)\%$ according to
\ull{the Monte} Carlo simulation by assuming \ull{a phase} space
distribution, where the errors are statistical due to limited
statistics of the Monte Carlo samples. Here in \ull{the
generator}, only \ull{the continuum} amplitude is considered.

\begin{figure}[htbp]
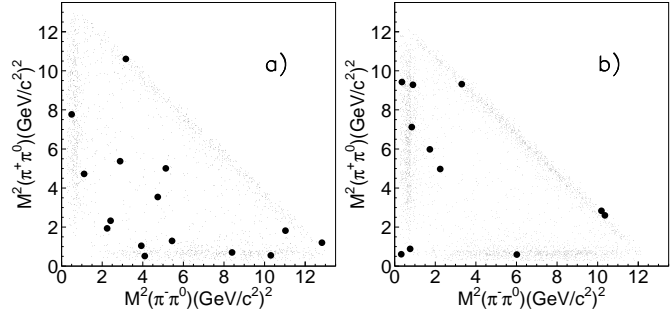

\centerline{ \hbox{\psfig{file=dalitz_pp3a.epsi,width=4.3cm}}
\hbox{\psfig{file=dalitz_c3b.epsi,width=4.3cm}}      }
\caption{Dalitz plots of $\EETO \threepi$ at (a) $\pspp$, and (b)
$\sqrt{s}=3.65$~GeV after the final selection.
The big black dots are for data, and the
small black dots are \ull{the Monte} Carlo simulated $\rhopi$ events. There
are still non-$\piz$ backgrounds in both \ul{plots for data.}} \label{dalitz}
\end{figure}

In order to evaluate the contributions of $\rhopi$ in the data
samples, the Dalitz plots of the $\threepi$ system are shown in
Figures~\ref{dalitz}a and \ref{dalitz}b
after requiring the invariant mass of the two photons lies between
0.10 and 0.17~GeV/$c^2$ (about $2\sigma$ around the $\piz$
nominal mass). The Monte Carlo predicted $\rhopi$ events are also
shown in the plots as small black dots. No clear $\rho(770)$ signal
can be seen either in \ull{the $\pspp$} or in \ull{the continuum}
data sample.

\begin{figure}[htbp]
\centerline{\hbox{\psfig{file=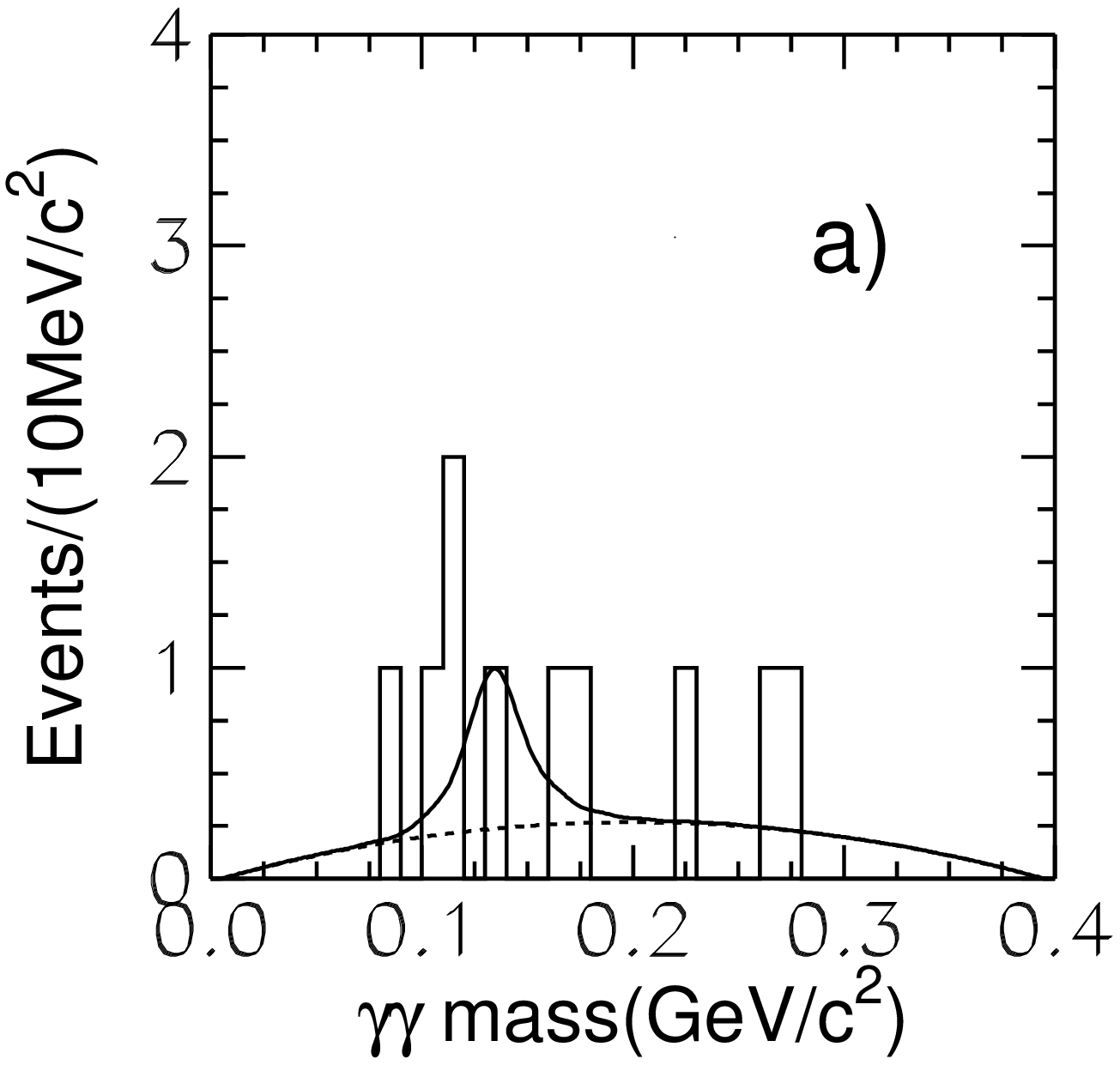,width=4.5cm}}
\hbox{\psfig{file=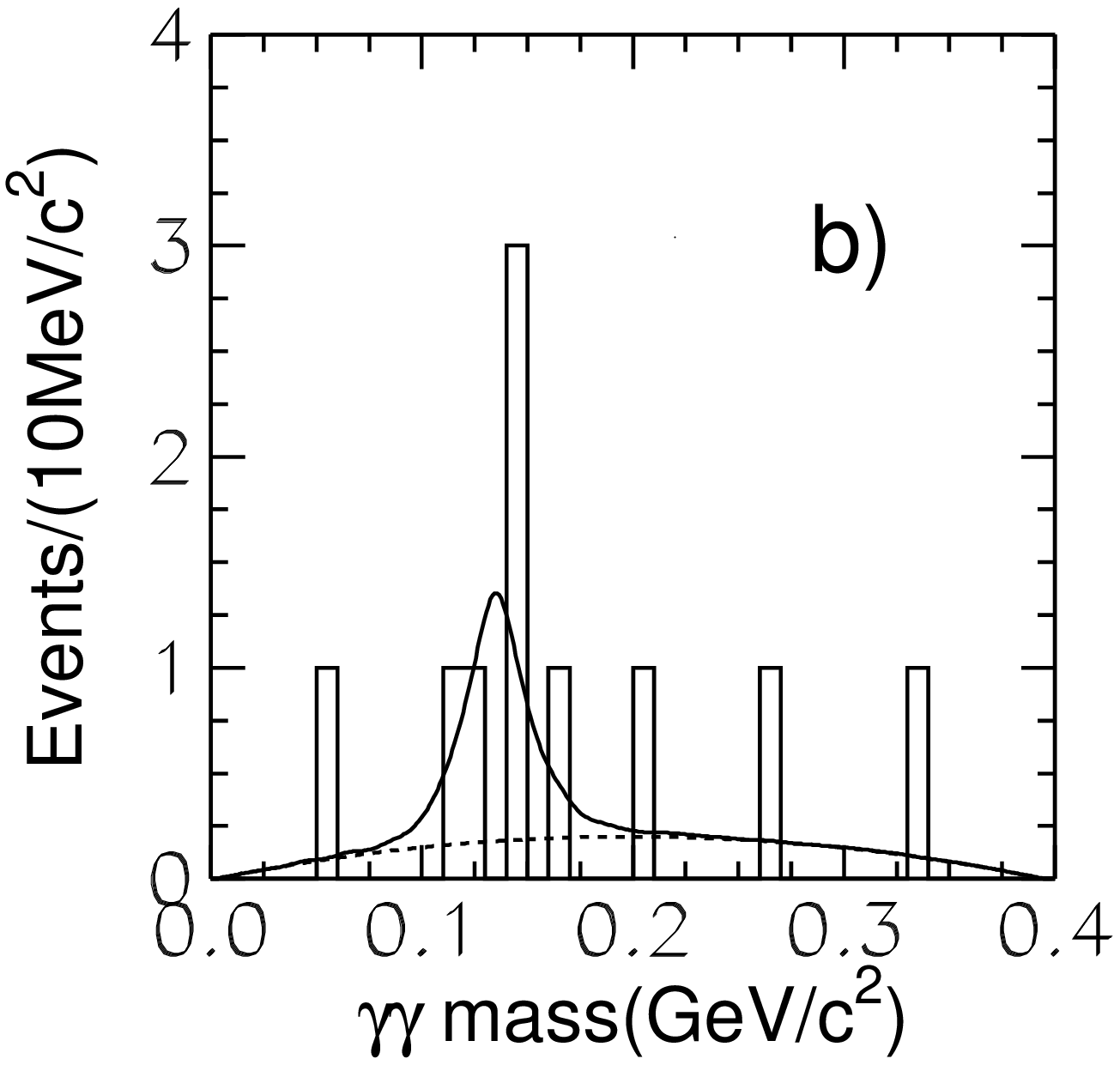,width=4.5cm}}   }
\caption{Invariant mass distributions of \ull{the two} photons of $\EETO
\rhopi$ at (a) $\pspp$ and (b) $\sqrt{s}=3.65$~GeV after making
the $\rho$ mass requirement. The histograms are data, and the
curves show the best fits.} \label{rhopi}
\end{figure}

Since the $\rhopi$ signals are not significant, we try to set the
upper limits for the $\EETO \rhopi$ cross sections at $\pspp$ and
at $\sqrt{s}=3.65$~GeV. This is done by fitting the two photon
invariant mass distributions (shown in Figure~\ref{rhopi}) after
requiring the $\pip\pim$ or $\gamma\gamma\pi^\pm$ invariant mass
in the range of 0.626 to $0.926~\gev/c^2$. The fits yield
$2.9\pm2.8$ and $4.9\pm2.7$ $\piz$s at $\pspp$ and
$\sqrt{s}=3.65$~GeV respectively. The upper limits at 90\% C. L.
on the numbers of events are 5.1 and 8.3 at $\pspp$ and
$\sqrt{s}=3.65$~GeV respectively. Here the systematic errors are
considered, see \ull{below}. The number of $\rhopi$ events are
overestimated since there are, in general, non-$\rhopi$
contributions to events containing $\piz$'s. The detection
efficiency is $(4.87 \pm 0.09)\%$ at $\pspp$ and $(5.14\pm
0.09)\%$ at $\sqrt{s}=3.65$~GeV from the Monte Carlo simulations
assuming \ull{a pure} continuum contribution.

\section{Systematic errors}

Many sources of systematic error are considered in the cross
section measurement. Systematic errors associated with the
efficiency are determined by comparing $\jpsi$ and $\psp$ data and
Monte Carlo simulation for very clean decay channels, such as
$\jpsi\ra\pip\pim\piz$, $\psp\ra\pip\pim\jpsi$, which allow the
determination of systematic errors associated with the MDC
tracking, trigger, kinematic fitting, photon identification
efficiency, \ul{requirement of photon number} and particle
identification~\cite{j3pi,pvt}. The uncertainty due to \ul{the
generator} for \ull{the $\threepi$} mode is estimated by the
efficiency difference between \ull{the phase} space and \ull{the
$\rhopi$} generators. The systematic errors from the
\ull{background estimation} are determined by comparing the
fitting result between the 2nd-order polynomial background and a
1st-order polynomial background and different fitting ranges in
the fits of \ull{the two-photon} invariant mass spectra.
Uncertainty of the integral luminosity of the data sample is also
a source of the systematic error.

All the sources considered are listed in Table~\ref{sys}. The
total systematic errors for $\EETO \rhopi$ are 20.8\% and 16.4\%
for \ull{the $\pspp$} and \ull{the continuum} data, respectively,
and for $\EE\ra\pip\pim\piz$  the corresponding numbers are 23.8\%
and 19.3\% for $\pspp$ and continuum data, respectively.

\btbl[htbp] \caption{\label{sys} Systematic errors for $\rhopi$
and $\threepi$ cross section measurements at $\pspp$ and
$\sqrt{s}=3.65$~GeV (\%).}
\begin{center}
\btbu{ l|c |c}\hline\hline Source   & $\sqrt{s}=3.65$~GeV
&$\pspp$\\ \hline\hline
Monte Carlo statistics        &  1.7 & 1.7\\
\hline Trigger &\multicolumn{2}{|c}{1.0}\\
\hline MDC tracking           &\multicolumn{2}{|c}{4.0}\\
\hline Kinematic fit          &\multicolumn{2}{|c}{6.0}\\
\hline Photon efficiency      & \multicolumn{2}{|c}{4.0}\\
\hline Number of photons      & \multicolumn{2}{|c}{2.0}\\
\hline Particle ID             & \multicolumn{2}{|c}{negligible}\\
\hline \ull{Background}         & 13.4    & 18.7 \\
\hline Luminosity             & 3.7     & 3.0   \\
\hline Generator (for $\threepi$ only)  & 10.2 & 11.5  \\
\hline\hline
Total $\left[\rhopi~(\threepi)\right]$ & 16.4~(19.3)& 20.8~(23.8)\\
\hline\hline
\etbu
\end{center}
\etbl

\section{Results and discussion}

We give the measurement of the cross section of $\EETO
\threepi$ and the upper limit of the cross section of $\EETO
\rhopi$ at $\pspp$ and $\sqrt{s}=3.65$~GeV. For the process
$e^+e^-\ra \threepi$, the Born order cross section is calculated with
\[
\sigma^B\left(\EETO \threepi\right)=\frac{N^{obs}}{\eff \cdot
\mathcal{L}\cdot \BR(\piz\ra\gamma\gamma)(1+\delta) },
\]
and for $e^+e^-\ra \rhopi$, the upper limit of the cross section
is calculated with
\[
\sigma^B\left(\EETO\rhopi\right)<\frac{N^{obs}_{UL}}{\eff \cdot
\mathcal{L}\cdot \BR(\piz\ra\gamma\gamma)(1+\delta)}.
\]

Here the initial state radiative correction factor, $1+\delta$,
and the efficiency, $\eff$, are obtained from the Monte Carlo
simulation assuming \ull{a pure} continuum contribution. Using the
numbers obtained from the above analysis (listed in
Table~\ref{cpnum}), one gets

\[\sigma^B(\EETO \threepi)=\ull{(8.4\pm 3.5 \pm 2.0)~pb},\]
\[\sigma^B(\EETO \rhopi)<6.0~pb\]
at $\pspp$ and
\[\sigma(\EETO \threepi)=(19.3\pm7.3 \pm 3.7)~pb,\]
\[\sigma(\EETO \rhopi)<25~pb\]
at $\sqrt{s}=3.65$~GeV, where the first errors are statistical
and the second ones are systematic, and the upper limits are at 90\% C. L.
\btbl[hbtp] \caption{\label{cpnum} Numbers used in the
calculations of $e^+e^-\ra \threepi$ and $\rhopi$ cross sections.}
\begin{center}
{\footnotesize \btbu{c| c |c | c |c } \hline\hline
               & \multicolumn{2}{|c}{$\sqrt{s}=3.65$~GeV}
               & \multicolumn{2}{|c}{$\pspp$}\\
\cline{2-5}
Decay Channel & $\threepi$ & $\rhopi$ & $\threepi$ & $\rhopi$ \\
\hline
 $N^{obs}(N^{obs}_{UL})$    & $10.0\pm 3.8$ & $4.9\pm
2.7~(8.3)$
                            & $11.4\pm4.7$ &$2.9\pm2.8~(5.1)$ \\
\hline
$\varepsilon(\%)$ & $7.88\pm 0.12$ &$5.14\pm0.09$ &$7.65\pm0.12$ & $4.87\pm0.09$\\
\hline
$1+\delta$  & 1.033 & 1.025  & 1.033 & 1.026     \\
\hline
$\mathcal{L}(pb^{-1})$
            & \multicolumn{2}{|c}{$6.42\pm0.24$}
            & \multicolumn{2}{|c}{$17.3\pm 0.5$}    \\
\hline
$\BR(\piz\ra\gamma\gamma)$~\cite{pdg}&\multicolumn{4}{|c}{$0.988$} \\
\hline\hline
\etbu
}
\end{center}
\etbl

The upper limit of $\EETO\rhopi$ cross section at
$\sqrt{s}=3.65~\gev$ is consistent with the measurement at
$\sqrt{s}=3.67~\gev$ by CLEO-c~\cite{cleocvp} and the calculation
from Ref.~\cite{wympspp}. The upper limit of $\sigma(\EETO
\rhopi)$ at $\pspp$ is a little lower than the upper limit of
$6.3~pb$ determined by Mark~III~\cite{mk3} and is lower than the
measurement of
\ul{$\sigma(\EETO\rhopi)=8.0^{+1.7}_{-1.4}\pm0.9~pb$ off the
$\pspp$ peak at $\sqrt{s}=3.67$~GeV by} CLEO-c~\cite{cleocvp},
which indicates that there must be a non-zero $\pspp\ra\rho\pi$
amplitude at the $\pspp$ energy.

\begin{figure}[htbp]
\centerline{
\hbox{\psfig{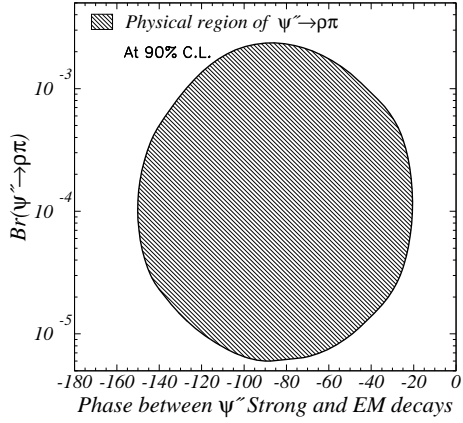}}}
\caption{Restriction on $\BR(\pspp\ra\rho\pi)$ and $\phi$ from the
measurement of $\rho\pi$ at \ull{the $\pspp$} peak in this
experiment. The hatched area indicate the physical region at 90\%
C. L.} \label{sborn}
\end{figure}

The measurement of $\sigma(\EETO \rhopi)$ at $\pspp$ also supports
the postulation in Ref.~\cite{wympspp} that the relative phase
between the strong and electromagnetic decays of $\pspp$ into
light hadrons is around $-90^{\circ}$. In this
scheme~\cite{wympspp}, the number of observed $\rho\pi$ events at
\ull{the $\pspp$} peak depends on both the $\pspp\ra\rho\pi$
branching fraction $\BR(\pspp\ra\rho\pi)$ and the relative phase,
$\phi$, between \ull{the $\pspp$} strong and electromagnetic decay
amplitudes. Using the measurement of $\rho\pi$ at \ull{the
$\pspp$} peak in this experiment, and the $\EE\ra\rho\pi$ cross
section at the continuum by the CLEO-c experiment~\cite{cleocvp},
\ull{a 2-dimensional} scan indicates that the physical region of
$\BR(\pspp\ra\rho\pi)$
 and $\phi$ is restricted
in the hatched area as shown in Figure~\ref{sborn} at 90\% C. L.
If the correlation between $\BR(\pspp\ra\rho\pi)$ and $\phi$ is
neglected, one gets
$\BR(\pspp\ra\rho\pi)\in(6.0\times10^{-6},~2.4\times10^{-3})$, and
$\phi\in(-150^\circ,~-20^\circ)$ at 90\% C. L.
This result is consistent with the calculation in
Refs.~\cite{rosnersd,wympspp} that the $\pspp \to \rhopi$
branching fraction is at the $10^{-4}$ level, and supports the
explanation of the ``$\rhopi$ puzzle'' observed between $\jpsi$
and $\psp$ decays by the $S$- and $D$-wave mixing model. We also
expect CLEO-c, with a few $fb^{-1}$ of $\pspp$ data~\cite{cleoc},
and BESIII, with even more $\pspp$ data~\cite{bes3}, to be able to
produce tighter constraints on $\BR(\pspp\ra\rho\pi)$ and $\phi$,
and give a better test of this scenario.

\section{Summary}

The processes $\EETO \threepi$ and $\rhopi$ are searched for at
$\pspp$ and $\sqrt{s}=3.65$~GeV. We observe $\EETO \threepi$
signals at the $3.1\sigma$ and $4.6\sigma$ levels for
$\sqrt{s}=3.773$~GeV and $\sqrt{s}=3.65$~GeV, respectively. No
significant $\rhopi$ signal is observed, and the upper limit of
the $\EETO \rhopi$ cross section at $\pspp$ is measured to be
$6.0~pb$ at 90\% C. L. assuming no $\pspp$ contribution.
Considering the interference between the continuum amplitude and
the $\pspp$ resonance amplitude, the branching fraction of $\pspp$
decays to $\rho\pi$ is determined to be
$\BR(\pspp\ra\rho\pi)\in(6.0\times10^{-6},~2.4\times10^{-3})$ at
90\% C. L.

\acknowledgments

   The BES collaboration thanks the staff of BEPC for their
hard efforts. This work is supported in part by the National
Natural Science Foundation of China under contracts
Nos. 10491303, 10225524, 10225525, the Chinese Academy
of Sciences under contract No. KJ 95T-03, the 100 Talents
Program of CAS under Contract Nos. U-11, U-24, U-25, and
the Knowledge Innovation Project of CAS under Contract
Nos. U-602, U-34(IHEP); by the National Natural Science
Foundation of China under Contract No. 10175060 (USTC),
and No. 10225522 (Tsinghua University).

\end{document}